\documentstyle[12pt]{article}
\begin{document}
\addtolength{\baselineskip}{.7mm}

\thispagestyle{empty}

\begin{flushright}
{\sc PUPT}-1484\\
hep-th/9407100  \\
July 1994
\end{flushright}
\vspace{.3cm}

\begin{center}
{\Large\bf{Phase Transition in Spherically Symmetric
 \\[2mm]   Gravitational Collapse of a Massless Scalar Field
}}\\[20mm]
{\sc Youngjai Kiem}\\[3mm]
{\it Joseph Henry Laboratories\\[2mm]
 Princeton University\\[2mm]
Princeton, NJ 08544\\[2mm]
E-mail: ykiem@puhep1.princeton.edu}
\\[30mm]

{\sc Abstract}
\end{center}

Phase transition in spherically symmetric collapse
of a massless scalar field is studied in 4-d
Einstein gravity. A class of exact solutions that
show the evolution of a constant incoming energy
flux turned on at a point in the past null infinity
are constructed to serve as an explicit example.
The recently discovered phase transition \cite{choptuik}
in this system is manifest; above a threshold value
of the incoming energy flux, a black hole is dynamically
formed and below that, the incoming flux is reflected
into the future null infinity. The critical exponent
is evaluated and discussed using the solutions.

\noindent

\vfill

\newpage

     Recent numerical investigations on gravitational
collapse of a
single massless scalar field in spherically symmetric Einstein
gravity revealed a remarkable critical behavior near the
onset of a black hole formation \cite{choptuik} \cite{evans}.
 A generic one-parameter class of solutions
$S[p]$ decomposes into two phases depending on the magnitude of $p$
that represents the strength of self-gravitational interaction
of incoming scalar field.
If $p<p^*$, where $p^*$ is a threshold value, the gravitational
collapse is followed by an explosion, reflecting back the incoming
stress-energy flux into the future null infinity.
For $p>p^*$, a black hold is formed and its mass shows
a universal scaling behavior $M_{BH}
\simeq  | p - p^* |^{\delta} $ near the threshold $p= p^*$.
These approaches are based on a formulation of Christodoulou
where this scattering process was studied as a Cauchy problem
in general relativity \cite{chris}.
In the context of 2-d dilaton gravity (CGHS model),
a simplified theoretical
model believed to capture many of the essential features of
4-d Einstein gravity, the explicit analytical understanding of
the similar phenomenon is possible
as the general (semi-)classical collapsing
solutions are available and well understood \cite{CGHS}.
To be specific, Strominger and Thorlacius found
a similar universal critical behavior
when Hawking radiation effect was taken into account
\cite{strominger}.
The explicit analytical understanding of this critical behavior in
4-d Einstein gravity, however, has not yet been complete, partly
due to the difficulty in solving 4-d Einstein gravity coupled with
a single massless scalar field in a dynamical scattering
situation.  For example, the rigorous proof of
 the universality and the
analytic calculation of the critical exponent remain to be studied.

     In this Letter, we make an attempt to analytically understand
this critical behavior.  As a first step, we show that the constant
incoming stress-energy flux of
scalar particles that was turned on at
some point in the past null infinity can not be reflected
forward into future null infinity
if the magnitude of the incoming flux
exceeds a certain threshold that plays a role of $p^*$.  Rather,
a black hole is dynamically formed.  Below this
threshold, the incoming flux implodes through the origin and
gets scattered forward into the future null infinity.
This analysis supports
the result of numerical studies, showing that this
one-parameter class of solutions decomposes
into the subcritical regime
and the supercritical regime.
The critical exponent that gives a scaling
relation between the order paramer, a black hole mass in this
case, and $|p - p^* |$ can be calculated for these solutions
in supercritical regime.
Our approach takes the classical back reaction of scalar matter
fields on the space-time geometry into account exactly, since this
critical phenomenon is the consequence of general relativity.
Therefore,
our result can not be observed in the leading order calculations
of the weak field approximation.  If we apply the weak field
approximation taking the Minkowskian metric as a fixed background,
any unbounded amount of incoming flux will be imploded
through the origin and all of it bounces forward into the
future.

     First, we derive a class of exact dynamical solutions that can
be utilized to construct an interesting physical situation.
Assuming spherical symmetry,
Einstein-Scalar action can be written as,
\begin{equation}
I= \int d^2 x \sqrt{-g} e^{- 2 \phi} ( R^{(2)} + 2 g^{\alpha \beta}
\partial_{\alpha} \phi \partial_{\beta} \phi - 2 e^{2 \phi}
- \frac{1}{2} g^{\alpha \beta} \partial_{\alpha} f \partial_{\beta} f
)
\label{action}
\end{equation}
where we integrated out the angular coordinates using Gauss-Bonnet
theorem.  Here our convention
is $(+---)$ metric signature and the 4-d metric is given by
$ds^2 = g_{\alpha \beta} dx^{\alpha} dx^{\beta} - e^{-2 \phi}
( d \theta^2 + \sin^2 \theta d \varphi^2 ) $.
$R^{(2)}$ and $f$ represent the scalar curvature of
the two dimensional
longitudinal metric $g_{\alpha \beta}$ and a massless
scalar field, respectively.  The gravitational constant $G$
is $16 \pi G = 1$ in our unit.  If we choose a conformal
gauge for the longitudinal metric $g_{\alpha \beta}$ as
$g_{\alpha \beta} = - e^{2 \rho + \phi} dx^+ dx^- $, the action,
Eq.(\ref{action}), reduces to
\begin{equation}
I = \int dx^+ dx^- ( 4 \Omega \partial_+ \partial_- \rho
- \frac{e^{2 \rho}}{\Omega^{1/2}} + \Omega \partial_+ f \partial_- f )
\label{conaction}
\end{equation}
along with two gauge constraints
\[  \partial_{\pm}^2 \Omega - 2
\partial_{\pm} \rho \partial_{\pm} \Omega
+ \frac{1}{2} \Omega ( \partial_{\pm} f )^2 = 0  \]
where $\Omega = e^{- 2 \phi }$.  The general static solutions of
the field equations obtained from the action (\ref{conaction})
under the gauge constraints can be readily found as follows;
we can consistently
reduce those partial differential equations into
the coupled second order ordinary
differential equations (ODE's) by assuming all functions depends on
a single space-like coordinate $x = x^+ x^- $.
The resulting ODE's can also
be derived from an effective action
\begin{equation}
I = \int dx ( x \dot{\Omega} \dot{\rho}+ \frac{1}{4}
\frac{e^{2 \rho}}{\Omega^{1/2}} - \frac{1}{4} \Omega x
\dot{f}^2  )
\label{xaction}
\end{equation}
and the original gauge constraints become
\[ \ddot{\Omega} - 2 \dot{\rho} \dot{\Omega }
+ \frac{1}{2} \Omega \dot{f}^2 = 0   , \]
where the dot represents taking a derivative with respect to $x$.
The complete solutions to this effective
action represent the general
solutions of the static Einstein-Scalar fields under a particular
choice of the conformal coordinates.
To obtain these, we observe that the action (\ref{xaction})
has three rigid continuous
symmetries that allow us to construct three corresponding
Noether charges, reducing the order of ODE's
by one.  The first symmetry is $f \rightarrow f + \alpha$, which
is clear as $f$ field appears only through its derivative.  The
second symmetry is $x \rightarrow x e^{\alpha}$ and
$ \rho \rightarrow \rho -  \alpha /2 $, which corresponds to the
translation of the asymptotically flat spatial coordinate.
The last symmetry transformation,
$ x \rightarrow x^{1 + \alpha}$, $\rho \rightarrow \rho
-  \ln ( 1 + \alpha )/4  - \alpha \ln x /2 $,
and $\Omega \rightarrow \Omega ( 1 + \alpha ) $, changes the
action by a total derivative.  This symmetry corresponds to the
rescaling of the asymptotically flat spatial coordinate.
The Noether charges for these symmetries  are constructed
to be
\begin{equation}
f_0 = x \Omega \dot{f} ,
\label{f0}
\end{equation}
\begin{equation}
c_0 = x^2 \dot{\rho} \dot{\Omega } + \frac{1}{2} x \dot{\Omega}
- \frac{1}{4} \Omega x^2 \dot{f}^2 - \frac{1}{4} x \frac{e^{2 \rho} }
{\Omega^{1/2}} ,
\label{c_0}
\end{equation}
and
\begin{equation}
x \dot{\rho} \Omega - \frac{1}{4} x \dot{\Omega} + \frac{1}{2}
\Omega = c_0 \ln x + c_1 .
\label{c_1}
\end{equation}
As these charges are conserved, $f_0$, $c_0$ and $c_1$ are
constants of integration independent of $x$.  The gauge constraint,
when combined with equations of motion from Eq.(\ref{xaction}),
reduces to a condition $c_0 = 0$.  We note that Eq.(\ref{c_1}) can be
readily solved for $\rho$ and by putting this into Eq.(\ref{c_0})
we get a decoupled ODE. Once $\Omega$ is solved in terms of $x$,
$\rho$ and $f$ are easily
determined as a function of $x$.  If we take
$f_0 = 0$, we recover static black hole solutions in Kruskal-Szeckers
coordinates with $c_1 =  4M^2$ where $M$ is the mass of a black hole.
For arbitrary value of $f_0$, $c_1 > 0$ case solutions are identical
to the solutions obtained by Janis {\em et.al.} using a technique to
generate some Einstein-Scalar solutions from the vacuum solutions of
the Einstein equations \cite{janis}.
Since $c_1 = 0$ case contains no black hole as long as the
static analysis is concerned and since the scalar field diverges
logarithmically near the horizon in $c_1 >0$ cases, our calculation
here is effectively a proof of no-scalar-hair theorem. The details
of this consideration can be found in \cite{kiem} where complete
static solutions of the more general action than Eq.(\ref{action})
are obtained.

     The $c_1 = 0$ cases are of particular importance
as we can generalize it to dynamical situations we are interested
in.  The static solutions in this case are calculated to be
\begin{equation}
\Omega = \frac{e^{-2 \rho_0}}{4}  ( e^{4 \rho_0}(\ln (x/x_0 ) )^2 -
4 f_0^2 ) ,
\label{pointstatic}
\end{equation}
\[ \rho = \frac{1}{4} \ln \Omega - \frac{1}{2} \ln x  +\rho_0 , \]
\[
f = \ln ( \frac{ \sqrt{e^{2 \rho_0} \Omega + f_0^2} - f_0 }
{ \sqrt{e^{2 \rho_0} \Omega + f_0^2 } + f_0 } )  + f_1
= \ln ( \frac{ e^{2 \rho_0} \ln  (x/x_0 ) -2 f_0 }
{ e^{2 \rho_0 } \ln (x/x_0 ) + 2 f_0 } )
+ f_1  , \]
where $\rho_0$, $f_1$ and $x_0$ are arbitrary constants.
For simplicity, we take $x_0 =1$, $\rho_0 = 0$ and $f_1 = 0$
for further discussions.  Then,
as $\Omega \rightarrow \infty$, the behavior of $f$ asymptotically
approaches to $- 2f_0 /r $
where the geometric radius $r$ is defined to be  $\sqrt{\Omega}$.
In this limit,
we find $r \rightarrow ( \ln x^+ +
\ln x^- ) / 2 $ and the 4-d metric becomes $ ds^2
\rightarrow  -
dx^+ dx^-/(x^+ x^- ) -
r^2 (d\theta^2 + \sin^2 \theta d\varphi^2 ) $.  After
a conformal transformation $\ln x^{\pm} \rightarrow x^{\pm}$, we
find that the asymptotic space-time is a flat Minkowskian.
Additionally, taking $G \rightarrow 0$ limit reproduces
the same results.
This limit is
the same as the weak field approximation
taking a Minkowskian metric as a fixed background metric.  The
wave equation for $f$ in $s$-wave sector
under this fixed background geometry has general
solutions of the form $f = - 2f_0 /r + constant$. Thus,
Eqs.(\ref{pointstatic}) are the relativistic generalization of
the point scalar charge solution. In a sense,
$c_1 =0$ limit is  similar to taking an extremal
limit of Reissner-Nordstr\" om black hole.
This consideration suggests
that the scalar charge $f_0$, in some dynamic situations,
can be a chiral field
instead of being a strict constant, for $f_0$ is indeed an arbitrary
chiral field in the weak field approximation. In general relativistic
case, however, this simple generalization is not possible
in general, since there can be a non-trivial corrections to
Eqs.(\ref{pointstatic}) of the order of $\partial_{\pm} f_0$.
In case of asymptotically steady incoming and outgoing stress-energy
flux, though, that simple generalization is
possible.  Forgetting about
global boundary conditions, the result of this extension is
\begin{equation}
\Omega = \frac{1}{4}
( \ln x )^2 - ( k_+ \ln x^+ - k_- \ln x^-  + q_0 )^2 ,
\label{scattering}
\end{equation}
\[ \rho = \frac{1}{4} \ln \Omega - \frac{1}{2} \ln x
 + \frac{1}{2} \ln ( 1 + 4 k_+ k_- ) ,  \]
\[ f = \ln ( \frac{ \ln x - 2 (k_+ \ln x^+ - k_- \ln x^- + q_0 ) }
                  { \ln x +
                    2 (k_+ \ln x^+ - k_- \ln x^- + q_0 ) } ) , \]
where $k_{\pm}$ and $q_0$ are constants.
We can straightforwardly verify that
these solutions satisfy field
equations derived from Eq.(\ref{conaction}) and the corresponding
gauge constraints.  The additional constant term in the expression
for $\rho$ is the correction term originating from $\partial_{\pm} f$.
This term turns out to be rather simple in this case where
the charge $f_0$ has terms only up to linear terms in $\ln x^{\pm}$.
The asymptotic stress-energy tensor averaged over the
transversal sphere
in a conformal coordinate system
that becomes asymptotically flat near the past or future infinity
is calculated to be $T_{\pm \pm} =  4  k_{\pm}^2$.  Thus,
$4k_{\pm}^2$ are interpreted to be an incoming and an outgoing energy
flux, respectively.  The similar asymptotic analysis shows that
$q_0$ represents the background component of the scalar charge.
Recently, $q_0 = 0$ and
$k_+ k_- = 0$ case of Eqs.(\ref{scattering}) was reported in
mathematics literature \cite{dc} and named scale invariant solutions.
For scattering situations in our consideration where incoming and
outgoing flux can coexist, a slightly
generalized version (\ref{scattering}) proves
to be useful.  Additionally,
the presence of $q_0$ term enables us to consider the time evolution
of the multiple square-type incoming
energy pulses by successively gluing
our solutions, unless a black hole
is formed in an intermediate stage.
To name a few other applications possible with our result, we
can construct
various scattering solutions, cosmological solutions
and point particle solutions
with time-varying charge at the origin, depending on the boundary
conditions and initial conditions.

     Using the results obtained so far, we
construct the one-parameter class of
exact solutions that reduce to $f = (2k \ln x^+ H ( \ln x^+ )
- 2k \ln x^- H ( - \ln x^- ) ) / r$ as we take the leading order
weak field approximation, i.e., taking $G \rightarrow 0$
limit.  Here, $H(x)$ denotes the usual step function.
This approximate solution in a {\it fixed} Minkowskian
background has a property that any
unbounded amount of the incoming stress-energy flux can be
totally reflected off from the origin into the future
null infinity.  However,
this picture qualitatively changes as we consider
the exact solutions focusing on the space-time geometry change
due to the stress-energy of the scalar field.
The physical region of space-time in our consideration is
specified by the
requirement $\Omega  \ge 0$, since the angular coordinates should not
have time-like signature.  Thus, the natural boundary is
$\Omega = 0$.
The asymptotic incoming wave section of the weak field
solutions is chosen as initial data and it represents
the turn-on of the constant incoming energy flux at a point
in past null infinity.  Under these conditions, the
class of solutions
parameterized by a constant $k \geq 0$ are found to be
\begin{eqnarray}
 \Omega =  \frac{1}{4}(u+v)^2               & \mbox{I} \nonumber \\
 \Omega =  \frac{1}{4}(u+v)^2 - k^2 v^2     & \mbox{II} \\
 \Omega =  \frac{1}{4}(u+v)^2 - (kv - k_- (k) u)^2 & \mbox{III} \nonumber
\label{ssol}
\end{eqnarray}
where we introduced $v = \ln x^+$ and $u = \ln x^-$.  Other
functions can be read off from Eqs.(\ref{scattering}).
The specific form of $k_- (k)$ depends on the boundary condition
at $\Omega =0$ in region III.  In our case, imposing a
covariant version of the reflecting boundary condition
yields $k_- (k) = ( \sqrt{1 + 8k -16k^2 } -1 ) /4 $.
The region I, bounded by the past null infinity, $v=0$ and
the origin $u = -v$, represents the Minkowski space
before the turn-on of the constant incoming flux.  The region
II, bounded by the future null infinity, $u=0$ and the past
null infinity, represents the propagation of the incoming
particles before any of them hits the boundary at the origin.
Note that our coordinate choice sets the turn-on time of the
incoming energy flux as $v = 0$.
Finally, the region III, bounded by $u=0$, the future null
infinity and the origin $u = -v (1-2k)/(1+2k_- (k) ) $, contains
the scattering of the incoming particles off the origin and
their further propagation toward the future null infinity.
As the matter particles hit the origin, they
linearly tilt it, adding some space-like component
to its tangent line.
If $k < 1/2$, the path of the origin remains time-like and
the incoming particles are scattered forward into the future
null infinity, just as in the subcritical regime of the
numerical simulations.  Especially, if $k \ll 1/2$, the weak field
approximation results are recovered and the geometry is almost
Minkowskian.  At $k = 1/2$, which can be interpreted as
a phase transition point, i.e., the black hole formation
threshold, the path of the origin, $u=0$, becomes
light-like for $v >0$.  If $k > 1/2$, the path of the origin becomes
space-like and form a trapped region in space-time.  In this
case, the region III disappears and our solutions here become
exactly the same as the one obtained in \cite{dc}.
As explained in detail in \cite{dc}, the resulting space-time
is a black hole with indefinitely increasing mass, for we
do not turn off the constant incoming flux.  Thus, this
corresponds to the supercritical phase of this scattering system.
As the numerical studies and the above considerations
suggest, the order parameter of this system is the black
hole mass.  Then, the important physical
quantity to compute, given our
exact one-parameter class of
solutions in supercritical regime, is the critical
exponent.
The geometric radius $r= r_H (v) $ of the apparent
horizon of the dynamic black hole in supercritical case,
determined by
$\partial_v r = 0$, is calculated to be
\begin{equation}
 r_H (v)= 2 (k- \frac{1}{2} )^{1/2} (k+ \frac{1}{2} )^{1/2} k v .
\label{mass}
\end{equation}
The $1/(2G) = 8 \pi$ times the value of this
corresponds to the apparent mass $M_A$
of the dynamic black hole.  The linear dependence on $v$ is
understandable as it is the time duration between the
turn-on of the incoming flux and the reference time $v$.
We also note that the angular averaged incoming energy
flux is $4k^2$.
Defining the transition point $p^* = 1/2$ and $p = k$, we find
that the critical exponent in this case is $\delta = 1/2$
in a scaling relation $M_A \simeq (p - p^* )^{\delta}$.

     In the aforementioned numerical study \cite{choptuik}, the
incoming flux is a pulse type and because of this, the
asymptotic out-region is a static black hole with
asymptotically flat geometry.
However, in our case, since we put infinite mass into the black
hole, the whole universe collapses to a future singularity, barring
the existence of any realistic out-region.  As a result, we should
consider turning off the incoming flux at a finite
time $v = v_1 > 0$.
Gluing a space-time with purely outgoing particles above $v = v_1$
is straightforward in subcritical case using
Eqs.(\ref{scattering}).
In supercritical case,
however, the gluing is highly non-trivial, for we can not directly
glue the static Schwarzschild solution at $v= v_1$.
For example, in
4-d Einstein gravity, gluing the Minkowski space to the Schwarzschild
geometry requires an impulsive shock-wave type injection of
matter particles \cite{dray}.  Thus, after
we turn off the source, the geometry goes through a brief transient
period to asymptotically settle down into a quasi-Schwarzschild
geometry.  During this process, the apparent horizon that was
initially space-like at the turn-off time
will settle down to a future null direction,
slightly changing its geometric radius.  The numerically
obtained $M_{BH} \simeq (p - p^* )^{\delta}$ with $\delta
\simeq 0.37$ is very difficult to calculate, as the scaling
relation
between $M_{BH}$ and $M_A$ gets complicated through this
process.  The similar situation
in the CGHS model is not as difficult as ours.
The complicated transient
behavior is absent in this case, due to the simplified
dynamics of the model.  Therefore, it
would be possible to directly glue a static
dilatonic black hole at $v=v_1$ and thereby getting a relation
$M_A \simeq M_{BH}$ if we had considered CGHS model from the
the outset.  An interesting observation in this regard is
that our
$\delta = 0.5$ is the same as the critical exponent obtained
by Strominger and Thorlacius in CGHS model \cite{strominger}.
There is
a reason for this connection as suggested in
\cite{birkhoff}.  The pure gravity sector of CGHS model,
other than being a target
space effective action from string theory,  can be considered
as a leading order theory in the $1/d$-expansion of the
spherically symmetric $d$-dimensional Einstein gravity.
We can, therefore, adopt $1/d$-expansion  and consider the leading
order behavior in the description of the
complex transient process in 4-d Einstein gravity.
As a zeroth order approximation, we glue a CGHS black hole
directly to our solutions to deduce
the approximate scaling relation $M_A  \simeq M_{BH}$.
Thus, the leading
order approximation of the exact critical exponent for  finite
pulse-type incoming energy flux is
now calculated to be $0.5$.
Since the next order correction to the critical exponent is
expected to be an order of $1/d = 0.25$ and the numerically
calculated value is about 0.37, our leading order value, $0.5$,
seems plausible.
Generalizing this consideration, it is conceivable
that the exact critical exponent for finite pulse-type incoming
energy flux  in
$d$-dimensional spherically symmetric Einstein gravity
will interpolate $\delta(4) \simeq 0.37 $ in 4-d Einstein gravity
and $\delta (\infty ) = 0.5$ in CGHS model
as a continuous function $\delta
(d) $ of the space-time dimensionality $d$.  In many other
cases of phase transitions, the critical exponent rapidly
saturates into a limiting value at $d = \infty$.
It will be an interesting
exercise to verify this conjecture and, additionally,
develop a systematic
perturbation theory with a dimensionless
expansion parameter $1/d$
to tackle other difficult problems in 4-dimensional gravity.

     The author wishes to thank H. Verlinde for useful
discussions.  After the completion of the preliminary version
of this manuscript, D. Christodoulou
pointed out that some results contained here were originally
discovered in \cite{dc} in a different context.
The author thanks him for that and other discussions.

\end{document}